# THE THEORETICAL SIMULATION OF OPTICAL PROPERTIES OF CRYSTALS WITH STATISTICALLY DISORDERED ARRANGEMENT OF ATOMS AND ATOMIC GROUPS ON THE BASIS OF POINT-DIPOLE MODEL.

**D.Yu. Popov *, O.A. Popova ****

*Institute of Chemistry FEBRAS, Vladivostok, Russia,

Email: popov@ich.dvo.ru

** Far Eastern Geological Institute FEBRAS, Vladivostok, Russia

With using of point-dipole model the theoretical calculations of main refractive indices and orientation of indicatrix of 18 minerals are performed. The feature of studied minerals is the statistically disordered arrangement of $CO_3$, $SO_4$, $SO_2$, $PO_4$ groups and also separate ions. The optical characters of uniaxial minerals and orientation of indicatrix of orthorhombic and monoclinic minerals, obtained by results of calculations, agree with experimental definitions.

The point-dipole model is enough widely used in investigations, which devoted to theoretical calculations of parameters of optical indicatrix of crystals and also to definition of polarizabilities of atoms [1-9]. For further development of this field the propagation of point-dipole model on a case of statistically disordered arrangement of atoms and atomic groups is obviously important. This was the purpose of the present work.

In terms of the point-dipole model, atoms in a crystal are considered as dipoles, whose dimensions are negligibly small in comparison with the interatomic distances. In this approximation, the local electric field induced by a light wave in the position *k* of the unit cell has the form [10]:

$$\mathbf{F}(k)=\mathbf{E}+4\pi \sum_{k'} \mathbf{L}(kk')\mathbf{P}(k')/v, \qquad (1)$$

where **E** – is the macroscopic field, **P** (*k* ') – is the dipole moment in the position *k* ', v – is the unit cell volume, **L** (*kk* ') – is the Lorentz-factor tensor, which depends on the geometry of the structure.

The dipole moment in the *k'* position is related to the local electric field **F**(*k*') in the same position by the equation:

$$\mathbf{P}(k')=\alpha(k')\mathbf{F}(k'), \qquad (2)$$

where $\alpha(k')$ - polarizability. The general electrical dipole moment of the unit cell has the form:

$$\mathbf{P}=\sum_{k}\alpha(k)\mathbf{F}(k). \qquad (3)$$

Substituting Eq. (2) into Eq. (1), we obtain the system of linear equations with respect to the components of the **F** vector. Solving this system and substituting obtained **F** vectors into Eq. (3) we obtain the tensor relating the total dipole moment of the unit cell to the vector of macroscopic field. Dividing the components of this tensor into the unit cell volume, one obtains the dielectric susceptibility tensor and can pass to the dielectric constant tensor.

In the case of presence in crystal structure of atomic groups arranged statistically, the calculation of the dielectric constant tensor becomes more complicated. In the first place, because the polarizabilities of atoms of such groups in Eq. (3) must be multiplied on the corresponding occupancies. The same must be made in Eq. (2), except for a case, when the atom in position *k'* belongs to the same atomic group, as atom in position *k* at $k \neq k'$. Secondly, because the occupation by a group of one of its possible positions means absence of another groups, whose atoms would be placed on forbidden distances from atoms of the given group. Due to this fact from the local fields (1) in positions *k*, in which atoms relating to atomic groups are placed it is necessary to subtract fields, equal to values of fields of dipoles in the not held positions *k'*, in which atoms of another groups would be placed:

$$\mathbf{F}'(kk') = \frac{3(\mathbf{P}(k')\mathbf{R}(kk'))}{r(kk')^4}\,\frac{\mathbf{R}(kk')}{r(kk')} - \frac{\mathbf{P}(k')}{r(kk')^3}, \qquad (4)$$

where **R** (*kk '*) - vector with the beginning in a position *k'* and end in a position *k*, r(*kk'*) - length of vector **R**(*kk'*).

With the help of the described approach we executed theoretical calculations of optical properties of a series of minerals (tab.1), for which disordering of $CO_3$, $SO_4$, $SO_2$, $PO_4$ groups, and also separate ions is characteristic. In the letter case the same approach, as for atomic groups, is applicable, because separate atoms are particular cases of atomic groups.

In tab.2 for all of studied minerals the types of a statistically disordered groups and atoms, and also Dmax - maximum of forbidden distances between positions of atoms are shown. Except cancrinite(I), in all of studied minerals there are forbidden distances between positions in a cell, in which atoms of the mentioned above groups are placed.

To perform the calculations, we wrote a program entitled AnRef3. The input data are the parameters of an elementary parallelepiped, the fractional coordinates of all the atoms in the unit cell, and their polarizabilities. The Lorentz-factor tensor is calculated by the method described in [10]. If the calculated dielectric-constant tensor is not diagonal, the program reduces it to the principal axes; then, the principal refractive indices equal to square roots of the diagonal components are calculated.

The polarizabilities of ions depend on many factors, e.g., on the bond polarity. As a consequence, they are essentially different in different compounds but remain close to the ionic refractions of the corresponding chemical elements [28, 29], which were used in our calculations as ionic polarizabilities.

The results of calculations of optical properties of uniaxial minerals are shown in tab. 3, orthorhombic mineral rhomboclase, and also monoclinic amphiboles - in tab. 4. Shown in these tables the experimental values for line D (λ = 589mμ) were taken from following works: for uniaxial minerals from [30], for rhomboclase from [31], for amphiboles from [32].

Pay on itself attention some deviations of the calculated main refractive indices from experimentally defined. This is caused by differences of refractions, used in calculations, from true polarizabilities of ions in crystals. Especially strong differences of the calculated and measured refractive indices for plumbogummite and hinsdalite are explained, apparently, by high ionic refraction of cation $Pb^{2+}$. Even the minor deviation of nature of bond of the given cation from an only ionic type results in an essential decrease of a general polarizability.

More interesting for us, however, were theoretical calculations not of main refractive indices, but parameters describing an anisotropy of optical properties, since these parameters depend mainly on geometry of structure, which can be defined relatively precisely. The polarizabilities of ions essentially vary in different compounds [28, 29], however their influencing on an anisotropy of optical properties can be not considerable. That was shown as a result of calculations of orientation of indicatrix in monoclinic and triclinic minerals [6-8].

On a lot of parameters describing an anisotropy of optical properties, the conformity between outcomes of calculations and experimental definitions is revealed. For all uniaxial minerals the optical character, obtained as a result of calculations, has coincided with experimentally defined. For orthorhombic mineral rhomboclase as a result of calculations the true orientation of an indicatrix relatively to the crystallographic axes was obtained.

Studied by us monocline amphiboles are characterized by the similar structural motives and close parameters of a unit cell. These minerals also have the similar one another orientation of indicatrix. The axis Nm in crystals of all minerals coincides with two-fold axis. This fact agrees with the results of calculations. Lying in a plane of symmetry m the axis of indicatrix Ng is turned down from a positive direction of axis *a* to the positive direction of axis *c* on angles 71-105°. The similar calculated angles are somewhat more - 115-132°. As a result of calculations the directions of axis Ng deflected from experimental directions on rather large angles were obtained. For kaersutite and hastingsite these angles are much less then 45°. This means that the calculated direction Ng is closer to a true direction Ng, than to a true direction Np. For both types of pargasite the given angle is close to 45°. For edenite difficultly to determine difference between calculated and true directions Ng because of absence of precise experimental data; it lays within the limits from 40 up to 61°.

The disagreements between the calculated and true parameters are caused by limitation of applicability of point-dipole model for explanation of optical properties of crystals [4]. To the main causes of this limitation it is necessary to relate, at first, that in calculations the isotropic polarizabilities of ions were used, secondly, that the dipoles were considered as placed in centers of ions. External electronic orbitals, which strongly influence on a polarizability of ion, can have center which is not coincides with center of ion, which was determined as a result of X-ray diffraction study.

**References**.

Table 1. Studied minerals.

| № | Mineral | Compound | Structural determination | |
|---|---|---|---|---|
| | | | R-factor | Ref. |
| 1 | Cancrinite (I) | $Na_8(Al,SiO_4)_6(CO_3)(H_2O)_2$ | 0.03 | [11] |
| 2 | Cancrinite (II) | $Na_6(Si,AlO_4)_6Ca_{1.5}(CO_3)_{1.5}(H_2O)_2$ | 0.028 | [12] |
| 3 | Gaudefroyite | $Ca_4Mn_3(BO_3)_3(CO_3)O_3$ | 0.017 | [13] |
| 4 | Gainesite | $Na_{1.08}K_{0.83}Zr_2Be(PO_4)_4$ | 0.055 | [14] |
| 5 | Abenakiite | $Na_{25.28}(Ce_3Nd_2La)(SO_2)(SiO_3)_6(PO_4)_6(CO_3)_6$ | 0.031 | [15] |
| 6 | Vishnevite | $K_{0.5}Na_{0.76}(Si,AlO_4)(SO_4)_{0.13}(H_2O)_{0.33}$ | 0.063 | [16] |
| 7 | Davyne | $Na_{3.06}K_{2.6}Ca_2(Al,SiO_4)_6(SO_4)_{0.5}Cl_2$ | 0.048 | [17] |
| 8 | Sugilite | $Na_2KFe_{1.66}Al_{0.34}Li_3Si_{12}O_{30}$ | 0.017 | [18] |
| 9 | Plumbogummite | $PbAl_3(P_{0.95}As_{0.05}O_4)_2(OH)_5H_2O$ | 0.037 | [19] |
| 10 | Osumilite | $(K_{0.78}Na_{0.22})(Mg_{0.92}Fe_{0.92}Mn_{0.16})(Al_{2.63}Fe_{0.37})(Si_{10.2}Al_{1.8})O_{30}(H_2O)$ | 0.066 | [20] |
| 11 | Hinsdalite | $PbAl_3(P_{0.69}S_{0.31}O_4)_2((OH)_{5.62}(H_2O)_{0.38})$ | 0.030 | [19] |
| 12 | Milarite | $KNa_{0.19}Ca_2(Al_{0.81}Be_{2.19}Si_{12}O_{30})(H_2O)_{0.67}$ | 0.029 | [21] |
| 13 | Rhomboclase | $(H_5O_2)Fe(SO_4)_2(H_2O)_2$ | 0.030 | [22] |
| 14 | Kaersutite | $(Na_{0.53}K_{0.41})Ca_{2.06}(Mg_{3.01}Fe_{1.07}Mn_{0.01}Ti_{0.52}Al_{0.34})(Si_{5.87}Al_{2.13})O_{22}(O\,H)_2$ | 0.056 | [23] |
| 15 | Edenite | $K_{0.33}Na_{1.22}Ca_{1.65}Sr_{0.01}Mg_{3.74}Fe_{0.85}Mn_{0.04}Ti_{0.16}Si_{6.94}Al_{1.06}O_{22}F_2$ | 0.029 | [24] |
| 16 | Hastingsite | $Na_{0.8}K_{0.2}Ca_2Mg_{0.55}Fe_{4.45}Al_{1.68}Si_{6.32}O_{23}(O\,H)$ | 0.057 | [25] |
| 17 | Pargasite (I) | $(Na_{0.79}K_{0.02})(Na_{0.05}Ca_{1.76}Fe_{0.19})(Mg_{3.42}Fe_{0.63}Al_{0.93}Ti_{0.02})(Si_{6.2}Al_{1.8})O_{22}(O\,H)_2$ | 0.026 | [26] |
| 18 | Pargasite (II) | $(Na_{0.91}K_{0.01})(Ca_{1.77}Na_{0.03}Mg_{0.07}Fe_{0.13})(Mg_{3.73}Fe_{0.31}Al_{0.93}Cr_{0.03})Si_{6.12}Al_{1.88}O_{22}F_{0.04}(OH)_{1.96}$ | 0.016 | [27] |

Table 2. Types of statistically disordered atomic groups and atoms in studied minerals.

| Mineral | Types of groups and atoms | Dmax, Å |
|---|---|---|
| Cancrinite (I) | $CO_3$ | |
| | O | 0.89 |
| Cancrinite (II) | $CO_3$ | 1.23 (O…O)<br>1.33 (C…C) |
| | O | 0.84 |
| Gaudefroyite | $CO_3$ | 1.56 (O…O)<br>0.82 (C…C) |
| Gainesite | $PO_4$, K, Na, Be | 0.97 (P…P)<br>1.48 (O…K)<br>1.11 (K…K)<br>1.36 (K…Na)<br>0.86 (Na…Na)<br>1.56 (K…Be) |
| Abenakiite | $SO_2$ | 1.43 (O…O) |
| Vishnevite | $SO_4$, K, Na | 1.18 (S…O)<br>1.46 (O…O)<br>1.31 (O…K)<br>1.03 (K…Na) |
| | O | 1.38 (O…O) |
| Davyne | $SO_4$, K, Na | 1.50 (O…O)<br>1.36 (K…O)<br>0.94 (K…Na) |
| Sugilite | Na | 0.375 |
| Plumbogummite | Pb | 0.576 |
| Osumilite | O | 2.074 |
| Hinsdalite | Pb | 0.439 |
| Milarite | Ca | 0.174 |
| | Na, O | 1.119 |
| Rhomboclase | O | 0.720 |
| Kaersutite | Na, K | 0.745 |
| Edenite | Na, K | 1.145 |
| Hastingsite | Na, K | 0.892 |
| Pargasite (I) | Na, K | 0.970 |
| | Na, Ca, Fe | 0.366 |
| Pargasite (II) | Na, K | 1.108 |
| | Na, Ca, Fe, Mg | 0.375 |

Table 3. Results of theoretical calculations of optical properties of uniaxial minerals.

| Mineral | No, Ne | Ng-Np | Opt. character |
|---|---|---|---|
| Cancrinite (I) | 1.525(1.515)<br>1.500(1.496) | 0.025<br>(0.019) | - (-) |
| Cancrinite (II) | 1.572(1.515)<br>1.530(1.496) | 0.041<br>(0.019) | - (-) |
| Gaudefroyite | 1.86 (1.81)<br>1.92 (2.02) | 0.06<br>(0.21) | + (+) |
| Gainesite | 1.661 (1.618)<br>1.697 (1.630) | 0.036<br>(0.012) | + (+) |
| Abenakiite | 1.623 (1.589)<br>1.613 (1.586) | 0.011<br>(0.003) | - (-) |
| Vishnevite | 1.515 (1.499)<br>1.509 (1.493) | 0.006<br>(0.006) | - (-) |
| Davyne | 1.531 (1.518)<br>1.538 (1.521) | 0.007<br>(0.003) | + (+) |
| Sugilite | 1.667 (1.610)<br>1.666 (1.607) | 0.001<br>(0.003) | - (-) |
| Plumbogummite | 1.942 (1.653-1.680)<br>1.964 (1.675-1.698) | 0.022<br>(0.018-0.022) | + (+) |
| Osumilite | 1.636 (1.540-1.546)<br>1.641 (1.546-1.550) | 0.005<br>(0.004-0.006) | + (+) |
| Hinsdalite | 1.932 (1.688)<br>1.965 (1.697) | 0.033<br>(0.009) | + (+) |
| Milarite | 1.627 (1.553)<br>1.622 (1.549) | 0.005<br>(0.004) | - (-) |

Table 4. Results of theoretical calculations of optical properties of orthorhombic and monoclinic minerals.

| Mineral | Ng, Nm, Np | Ng-Np | Opt. character, 2V,° | ∠ *a* Ng, ° |
|---|---|---|---|---|
| Rhomboclase | 1.647 (1.635)<br>1.599 (1.550)<br>1.572 (1.533) | 0.076□(0.102) | + (+)<br>76 (27) | |
| Kaersutite | 1.680 (1.700 – 1.772)<br>1.649 (1.690 – 1.741)<br>1.648 (1.670 – 1.689) | 0.032<br>(0.019 – 0.083) | + (-)<br>23 (66 - 82) | 123 (86-<br>105) |
| Edenite | 1.685 (1.632 – 1.730)<br>1.677 (1.618 – 1.714)<br>1.668 (1.615 – 1.705) | 0.017<br>(0.014 – 0.026) | - (-, +)<br>84 (27 - 95)[*] | 132 (71-92) |
| Hastingsite | 1.735 (1.730)<br>1.735 (1.729)<br>1.695 (1.702) | 0.040<br>(0.028) | - (-)<br>8 (10) | 115 (93) |
| Pargasite (I) | 1.714 (1.635)<br>1.700 (1.618)<br>1.687 (1.613) | 0.026<br>(0.022) | + (+)<br>87 (60) | 123 (79) |
| Pargasite (II) | 1.709 (1.635)<br>1.695 (1.618)<br>1.685 (1.613) | 0.025<br>(0.022) | + (+)<br>83 (60) | 123 (79) |

*- 2V